\documentclass[a4paper,12pt]{article}

\usepackage[top=2.5cm,bottom=2.5cm,left=2.5cm,right=2.5cm]{geometry}

\usepackage{graphicx}
\usepackage{epstopdf}
\usepackage{subfigure}
\usepackage{amssymb,amsbsy,amsmath,amsfonts,epic,eepic,amsthm}
\usepackage{titling}
\usepackage[numbers]{natbib}
\usepackage{xcolor}
\usepackage{soul}

\begin{document}

\title{Reservoir Computing based on\\ 
Quenched Chaos}

\author{Jaesung~Choi~and~Pilwon~Kim
\thanks{The authors are with Department of Mathematical Sciences, Ulsan National Institute of Science and Technology(UNIST), Ulsan Metropolitan City, 44919,  Republic of Korea. e-mail: pwkim@unist.ac.kr.}}

\maketitle

\begin{abstract}

Reservoir computing(RC) is a brain-inspired computing framework that employs a transient dynamical 
system whose reaction to an input signal is transformed to a target output. One of the central problems in RC is to find a reliable reservoir with a large criticality, since computing performance of a reservoir is 
maximized near the phase transition. In this work, we propose a continuous reservoir that utilizes transient dynamics of coupled chaotic oscillators in a critical regime where sudden 
amplitude death occurs. This ``explosive death'' not only brings the system a large criticality which provides a 
variety of orbits for computing, but also stabilizes them which otherwise diverge soon in chaotic units. 
The proposed framework shows better results in tasks for signal reconstructions than RC based on 
explosive synchronization of regular phase oscillators. We also show that the information capacity of the 
reservoirs can be used as a predictive measure for computational capability of a 
reservoir at a critical point. 
\end{abstract}



\section{Introduction}
Recently, reservoir computing has emerged as a promising computational framework for utilizing a dynamical system for computation. While an input stream perturbs the transient intrinsic dynamics of a medium(``reservoir''), a readout layer is trained to extract features out of such perturbations to approximate a target output. Due to its complex high-dimensional dynamics, the reservoir serves as a vast repertoire of nonlinear transformations that can be exploited by the readout. The major advantage of reservoir computing is their simplicity in training process compared to other neural networks. Another advantage is their universality in that they can be realized using physical systems, substrates, and devices \cite{R1, GOUD, DU20}.

There is the hypothesis that a system can exhibit maximal computational power at a phase transition between ordered and chaotic behavioral regimes \cite{EC27, EC60}. It has been observed that the brain operates near a critical state in order to adapt to a great variety of inputs and maximize information capacity \cite{EC18, EC19, EC20}. Perturbations occurring in a critical regime neither spread nor die out too quickly, providing the most flexibility to the system \cite{EC21, EC22}. This concept of “computation at the edge of chaos” may also have an implication to material computation, whereby a material has the most exploitable properties \cite{EC75}. More extensive review on this subject can be found in \cite{EC1}.

In RC, designing a reservoir which has a large criticality is important to perform complex tasks. In case of a reservoir based on continuous dynamical systems, one can create criticality by tuning intrinsic parameters so that the reservoir operates at a bifurcation point across which the dimension of the attractor abruptly declines. We call such system a critical reservoir. A system of coupled oscillator exhibits a first order transition from incoherent state to synchronized state that occurs under a specific relation between the coupling strength and connectivity, which is called explosive synchronization. In the previous work \cite{CP}, we showed that a reservoir of coupled Kuramoto oscillators near explosive synchronization forms a critical reservoir and performs excellent computations.

Amplitude death(AD) is another way to create a criticality in coupled oscillatory units. It indicates complete cessation of oscillations induced from change in intrinsic parameters of the system. The occurrence of AD has been found in the case of chemical reactions \cite{AD5, AD7}, neuronal systems  \cite{AD1,AD2} and coupled laser systems \cite{AD11, AD12}. 
It has been also reported that AD can occur abruptly in a systems of a coupled nonlinear oscillators \cite{ED1,ED2,ED3,ED4}. Such simultaneous cessation of oscillations is the first-order transition to AD and called ``explosive death''(ED).  

In thiw work, we focus on computing ability of chaotic systems near a criticality created in the form of ED. There have been many researches on chaos computing \cite{RCC1,RCC2}, even in the context of RC \cite{RCC3,RCC4,RCC5,RCC6}; Chaos computing takes advantage of an infinite number of orbits/patterns inherent in the attractor to be used for particular computational tasks. It  also utilizes the sensitivity to initial conditions of chaotic systems to perform rapid switching between computational modes. However, chaos computing often has a control problem to stabilize particular orbit.

Our major goal is to construct a chaos based reservoir with a large criticality induced from explosive death. 
We use the coupled chaotic oscillators and adjust a coupling strength so that they remain near the stage of ED. A reservoir in such critical regime provides a large variety of orbits in transient dynamics which can be used in computational tasks. Different from previous chaotic computing methods, the reservoirs still remain in a regular regime during computation, but close enough to chaos to enjoy its richness. We also investigate how such quenched chaos enhances the criticality in RC. To show the contribution of chaos in criticality, we compare the computation performance of a chaotic critical reservoir with a nonchaotic critical reservoir.

\section{MODEL}

\subsection{Reservoir of nonidentical chaotic elements}

We consider the reservoir that consists of coupled chaotic oscillators. The reservoir is supposed to suppress chaotic oscillations in its ground state ready for external signals. Once designated oscillators are excited by inputs, the deviation of the oscillators from the ground state is closely observed until they return to the ground state. The basic idea underlying the oscillatory reservoir computing is that, if network is large enough, all the information necessary to construct proper computational results can be found in the transient trajectories aroused by inputs.

Recently, the occurrence of an explosive death transition has been found in chaotic oscillator coupled via mean–field diffusion\cite{ED3}. To extend this result to nonidentical oscillators, we consider a reservoir that consists of $N$ Lorenz systems coupled via a mean–field diffusion as,
\begin{equation} \label{eq1}
\begin{aligned}
	&\frac{1}{w_i}\frac{dx_i}{dt} = 10(y_i-x_i)+K(Q\bar{x}-x_i) \\	
	&\frac{1}{w_i}\frac{dy_i}{dt} = -x_iz_i+\rho x_i-y_i \\
	&\frac{1}{w_i}\frac{dz_i}{dt} = x_iy_i-\frac{8}{3}z_i \\
\end{aligned}
\end{equation}
where $i = 1,\cdots , N$ is the index of the oscillators and $\bar{x}=\frac{1}{N}\sum_{i=1}^{N}x_i$ is the mean field of the state variable $x$.  
The parameter $K$ is the strength of coupling and $0\leq Q\leq 1$, is the intensity of the mean field. Each single system exactly concides with the conventional Lorenz system if $K=0$ with $w_i=1$. Here we use $Q = 0.7$, following \cite{ED3}. If the frequencies of nodes are identical, then the systems reverts 
to the one in \cite{ED3}. When running the system in Eq. (1) as a reservoir, we set the parameters for the system to be posed in a critical regime where the phase transition occurs. The adjustment of the 
parameters according to an order parameter will be discussed in Section 3.

\subsection{Readout and training}

The chaotic reservoirs are applied to supervised tasks of which training data comes in the form of $(u,v)$ where $u(t) = (u^1(t), \cdots , u^p(t))\in \mathbb{R}^p$ is an input signal and $v(t) = (v^1(t), \cdots , v^q(t))\in \mathbb{R}^q$ is a target output. We assign $p$ nodes of the reservoir as input nodes. Before the training process starts, we run the network until it reaches an amplitude death state. Then the input stream $u(t)$ is fed to the reservoir, in a way that the value of $x(t)$ in the input nodes are perturbed by adding $u(t)$. All evolutionary activities of the nodes are measured to compute the output function $f_{\text{out}}=(f^1_{\text{out}},\cdots,f^q_{\text{out}})\in\mathbb{R}^q$.

In the readout process, it is better to use not only the past values of the nodes as well as the current 
ones,  to exploit the rich dynamics of the chaotic reservoirs. Here we use a output function that takes 
past $s$ sampled values of the frequency $x_i'=\frac{dx_i}{dt}$ at discrete times $t-\Delta t,t-2\Delta t,\cdots,t-s\Delta t$ and maps them to the desired output at time $t$. We define the output function $f_{\text{out}}=(f^1_{\text{out}},\dots,f^q_{\text{out}})\in\mathbb{R}^q$ of $(s,\Delta t)$-type as
\begin{equation} \label{eq2}
f^l_{\text{out}}(t) = \sum_{i=1}^{N}\sum_{j=1}^{s}w^l_{i,j}x_i'(t-(j-1)\Delta t), \quad l = 1,\cdots,q
\end{equation}
Here $w^l_{i,j}$ are weights to be determined from the training process for each computational task, so that $f_{\text{out}}(t)$ is as close to $v(t)$ as possible. For example, if the output data is a time series $v(t_1),v(t_2),\cdots,v(t_M)$, the mean-square error
\begin{equation} \label{eq3}
\frac{\sum_{i=1}^{M} \lVert v(t_i)-f_{\text{out}}(t_i)\rVert^2}{\sum_{i=1}^{M}\lVert v(t_i)\rVert^2}
\end{equation}
can be used to determine the weights $w^l_{i,j}$. Note that minimizing the error in Eq. (3) with respect to the weights $w^l_{i,j}$ in Eq. (2) corresponds to a linear least squares problem.

\section{Creating a criticality by the explosive death}

Benefit of using the coupled chaotic systems in Eq. (1) as a reservoir  is that one can easily create a 
large criticality with a first order phase transition in the system. Across a critical point of the coupling 
force, the compound oscillatory motions of the system collapse into an equilibrium point. This fixed 
equilibrium state near a critical point is used as the ground state for reservoir computing, where the 
system always returns to after every computation, erasing unnecessary information from previous evaluations and preparing for the next inputs.

To look for a possible phase transition in Eq. (1), we define an order parameter $r_{\text{var}}$ in terms of the variation of amplitudes, as
\begin{equation} \label{eq4}
r_{\text{var}} = \frac{1}{N}\sum_{j=1}^{N}\exp(-c\, \text{var}_j), \quad c>0
\end{equation}
where $\text{var}_j$ is the temporal variance of the frequency $x_j'(t)$ \cite{CP}. This is a measure for desynchrony that sensitively shows a degree of deviation of oscillators from a steady frequency. In the ground state, the temporal variance of the frequency should be kept low for reliable computations. Note that, for each 
oscillator, the temporal variance of the frequency becomes 0 if a strong coupling strength holds 
oscillators in a phase-locked state, keeping their common frequency steady.
We also use another order parameter based on the normalized average amplitude as
\begin{equation} \label{eq5}
A(K) = \frac{a(K)}{a(0)}, \quad a(K) = \frac{\sum_{i=1}^{N}(\langle x_{i,max}\rangle_t-\langle x_{i,min}\rangle_t)}{N},
\end{equation}
which is widely used for chaotic oscillators  \cite{ADO2,ADO1,ED3}. Note that $A(K)=0$ implies a complete cessation of oscillations and $A(K)=1$ imples nondepressed chaotic oscillations.

\begin{figure}[htb]
	\centering
	\begin{tabular}{@{}cccc@{}}
		\includegraphics[width=.35\textwidth]{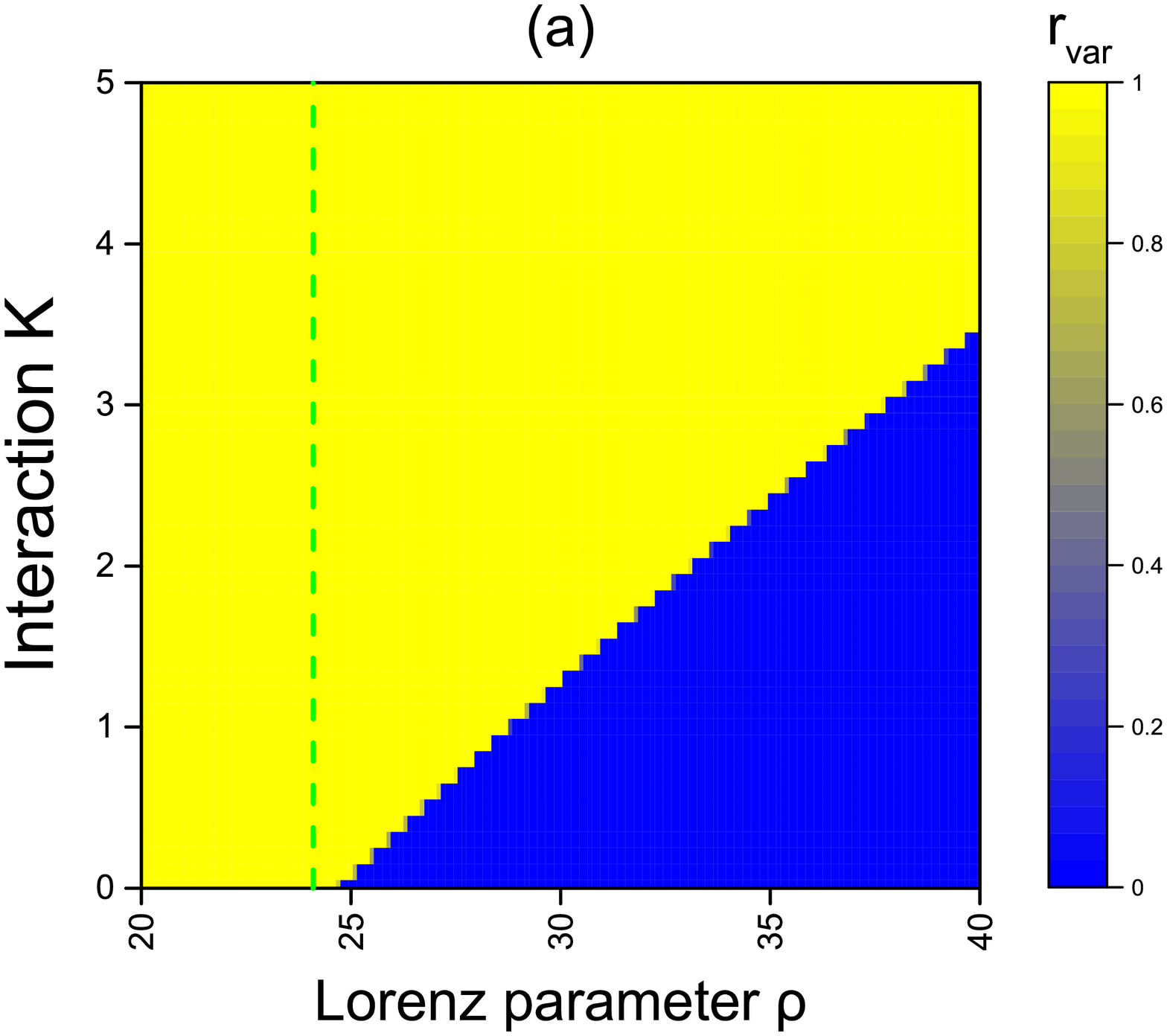} &
		\includegraphics[width=.35\textwidth]{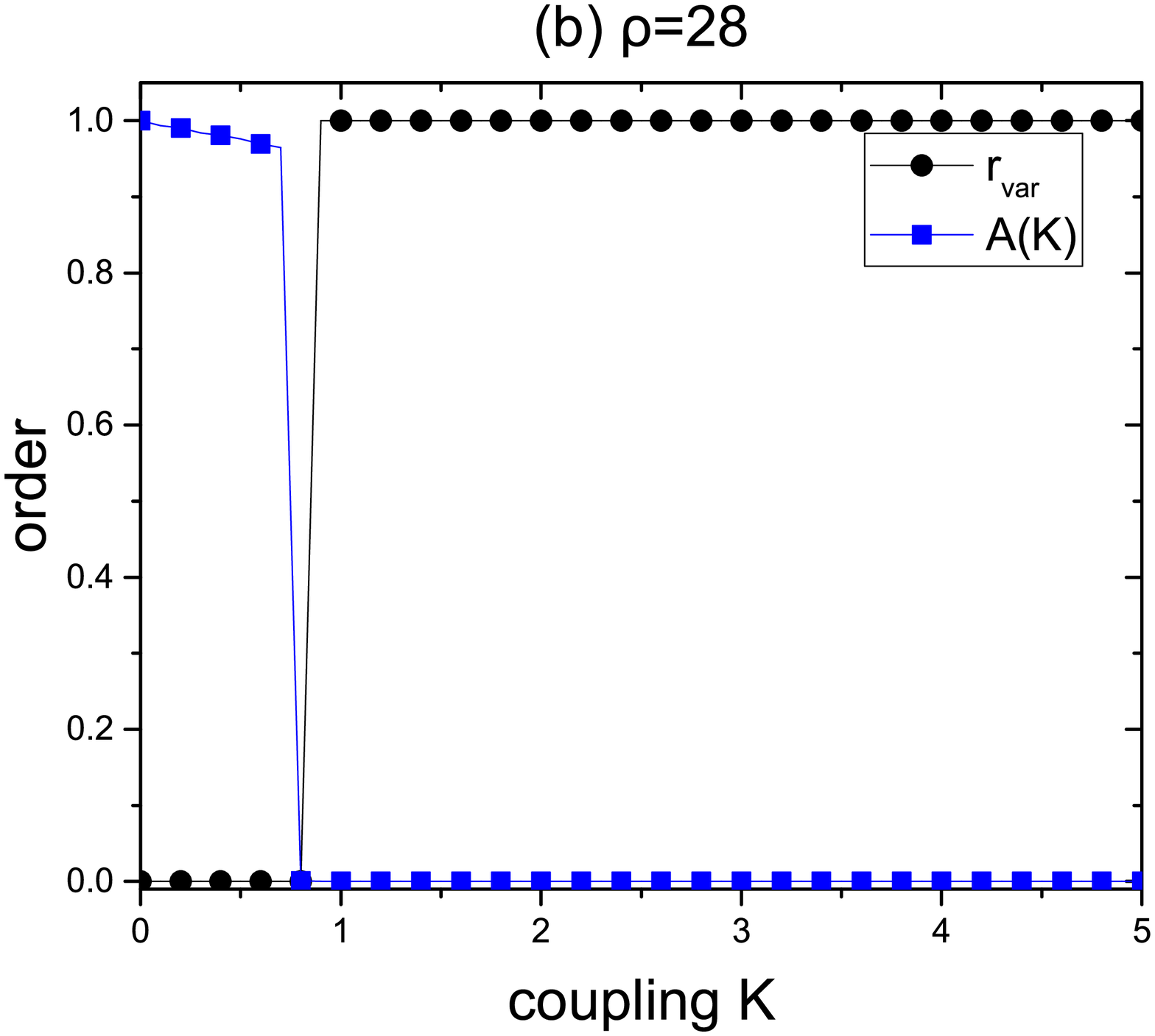} &
		\includegraphics[width=.35\textwidth]{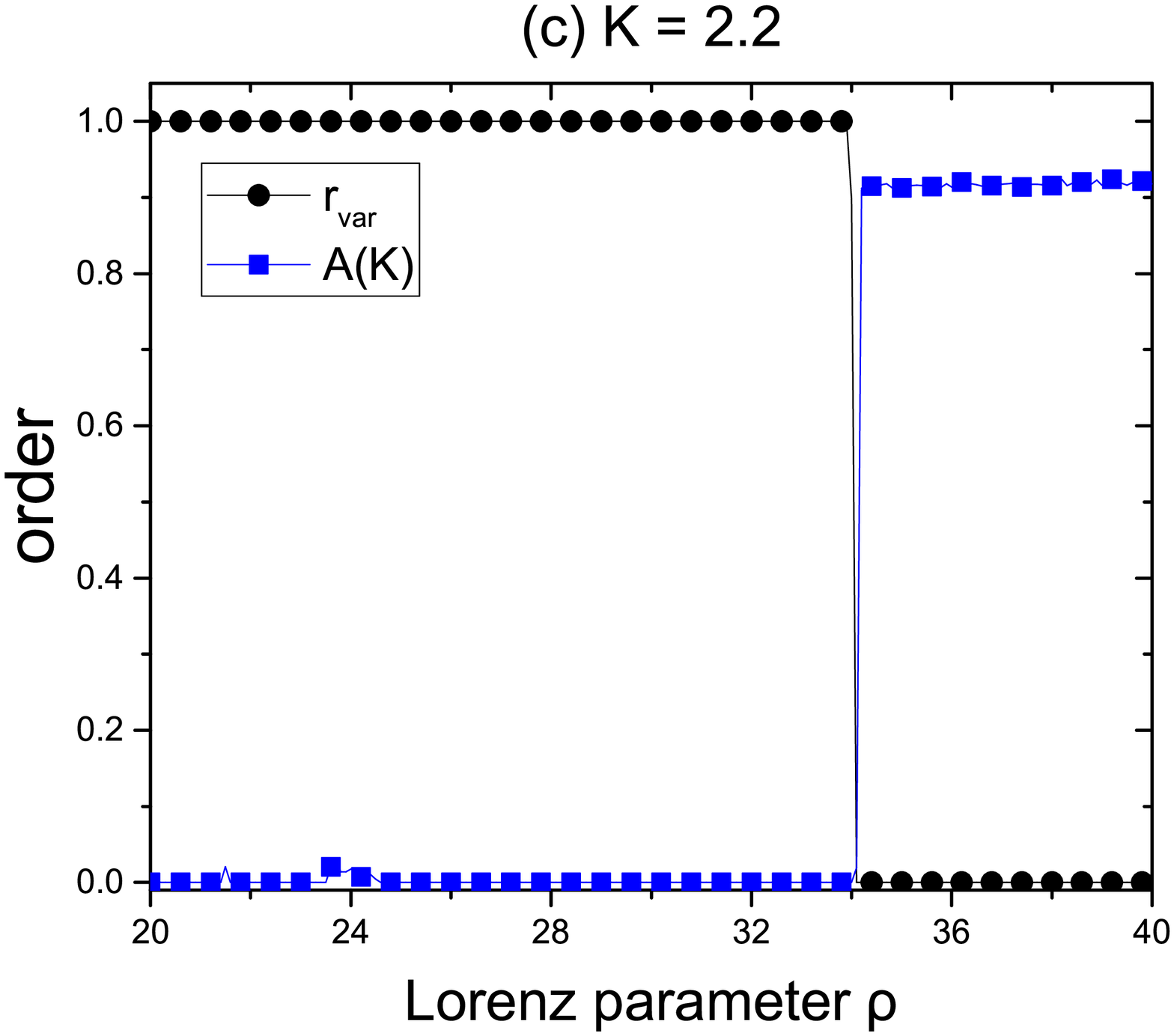} &
	\end{tabular}
	\caption{(a) The phase diagram of order parameter $r_\text{var}$ according to the coupling strength $K$ and lorenz parameter $\rho$. The dotted vertical line is drawn at $\rho=24$ to denote separation of non-chaotic dynamics(left) and chaotic dynamics(right) in the case $K=0$. (b) Cross-sectional figures of two order parameters, $r_\text{var}$ and $A$, according to $K$ with $\rho=28$ fixed.  (c) Cross-sectional figures of two order parameters, $r_\text{var}$ and $A$, according to $\rho$ with $K=2.2$ fixed.}
\end{figure}

If the frequencies of nodes in Eq. (1) are identical, then the system exhibits explosive death, the 
discontinuous transition from the oscillatory state to the completely quenched state \cite{ED3}. Indeed, Eq. (1) with nonidentical natural frequencies still exhibits the same phenomena. Figure 1(a) is the phase diagram of the order parameter $r_\text{var}$ in the $\rho$-$K$ parameter plane. We use $N=100$ oscillators whose natural frequencies follow the uniform distribution in $[1,1.3]$. The dynamical states of system is obtained by backward continuation from a large value of $K$ \cite{ED3}. The order parameter $r_\text{var}$ was averaged between $t=[3900, 4000]$. The diagonal line in Figure 1(a) clearly splits the parameter plane according to the value of $r_\text{var}$, indicating  that the discontinuous phase transition occurs with respect to both  $\rho$ and $K$. The graphs in Figure 1(b) and (c) indicate cross-sectional figures of two order parameters in $K$ and $\rho$ directions, respectively. It is verified that both of the order parameters $r_\text{var}$ and $A$ exhibit extremely abrupt jump at the same critical point. 
\section{Results}

\subsection{Numerical tests}
In the following numerical examples to test the learning ability of the oscillator networks, we use the $(10,0.1)-$type readout. That is, the output function $f_{\text{out}}$ at $t$ is obtained from $10$ previous sampled values of the oscillator frequencies $x_i'(t), \cdots, x_i'(t-0.9)$ in Eq. (2).

We set up two types of tasks, inferring missing variables and filtering signals, both of which require the 
presence of long-term memory for proper execution.  In the first type of tasks, RC is used to reconstruct values of hidden variables of the systems from observation of a single variable. For example, suppose a temporal 
data $(x(t),y(t),z(t))$ is generated from an unknown system of differential equations. The reservoir is trained to infer $y(t)$ or $z(t)$ from $x(t)$. This implies that RC implicitly learns a structure of the system that generates the corresponding data. We use the data generated from two chaotic systems, the Rössler system and the Chua's circuit. 

The filtering task is to learn the scalar output
\begin{equation} \label{eq6}
v(t) = \frac{1}{m}\sum_{k=1}^{m}\left( a u(t-k)+b u(t-k)^2+c u(t-k)^3 \right)
\end{equation}
which is determined from the past $m$ values of an input stream $u(t)$. Here $a,b$ and $c$ are some 
nonzero parameters. If $m=1$, the task is simply to implement a polynomial function of the current value of the input. The task becomes more challenging as $m$ increases, requiring long-term memory to evaluate averaged values. In our tasks, we use the parameters $m=20, a=1, b=3$ and $c=5$, and have the input $u(t)$ generated from the Mackey-Glass equation which provides standard benchmark task for chaotic series handling \cite{TK}. 

In each task, the continuous input signal $u(t)$ and the target signal $v(t)$ are generated for $t \in [0,6000]$.
To make sure that the system is positioned in the reliable ground  state, we skip first 1000 time steps of the output. The training process is applied to match $f_{\text{out}}$ to $v(t)$ over the 4,000 discrete time steps, $t=1001,1002,\cdots,5000$. That is, the readout weights $w^l_{i,j}$ in Eq. (2) are determined to minimize the relative error in Eq. (3). Then we evaluate the relative error between $f_{\text{out}}$ to $v(t)$ as the performance measure over 1,000 discrete sampled time steps  for $t \in (5000,6000].$ 

\begin{figure}[htb]
	\centering
	\begin{tabular}{@{}cccc@{}}
		\includegraphics[width=.35\textwidth]{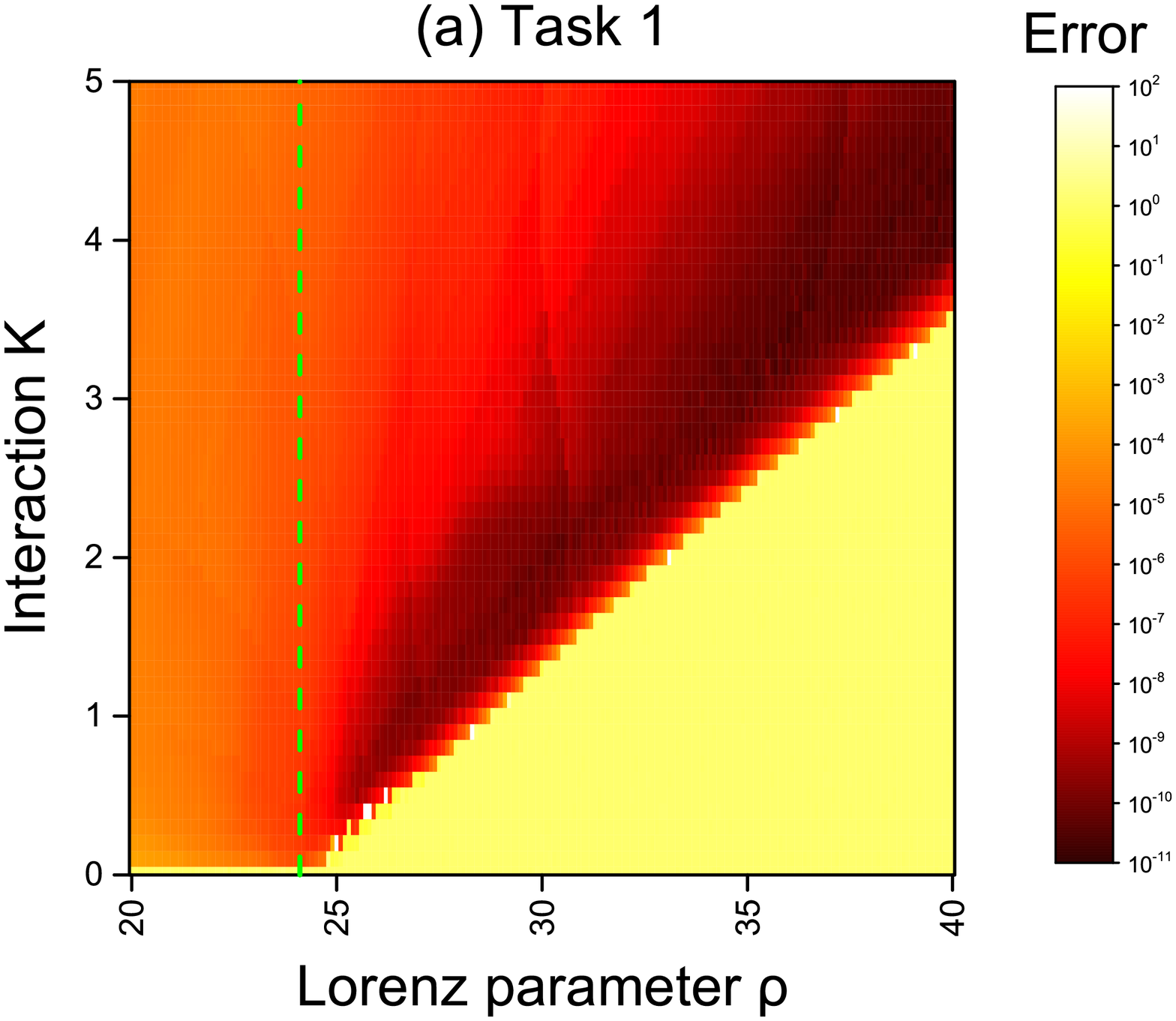} &
		\includegraphics[width=.35\textwidth]{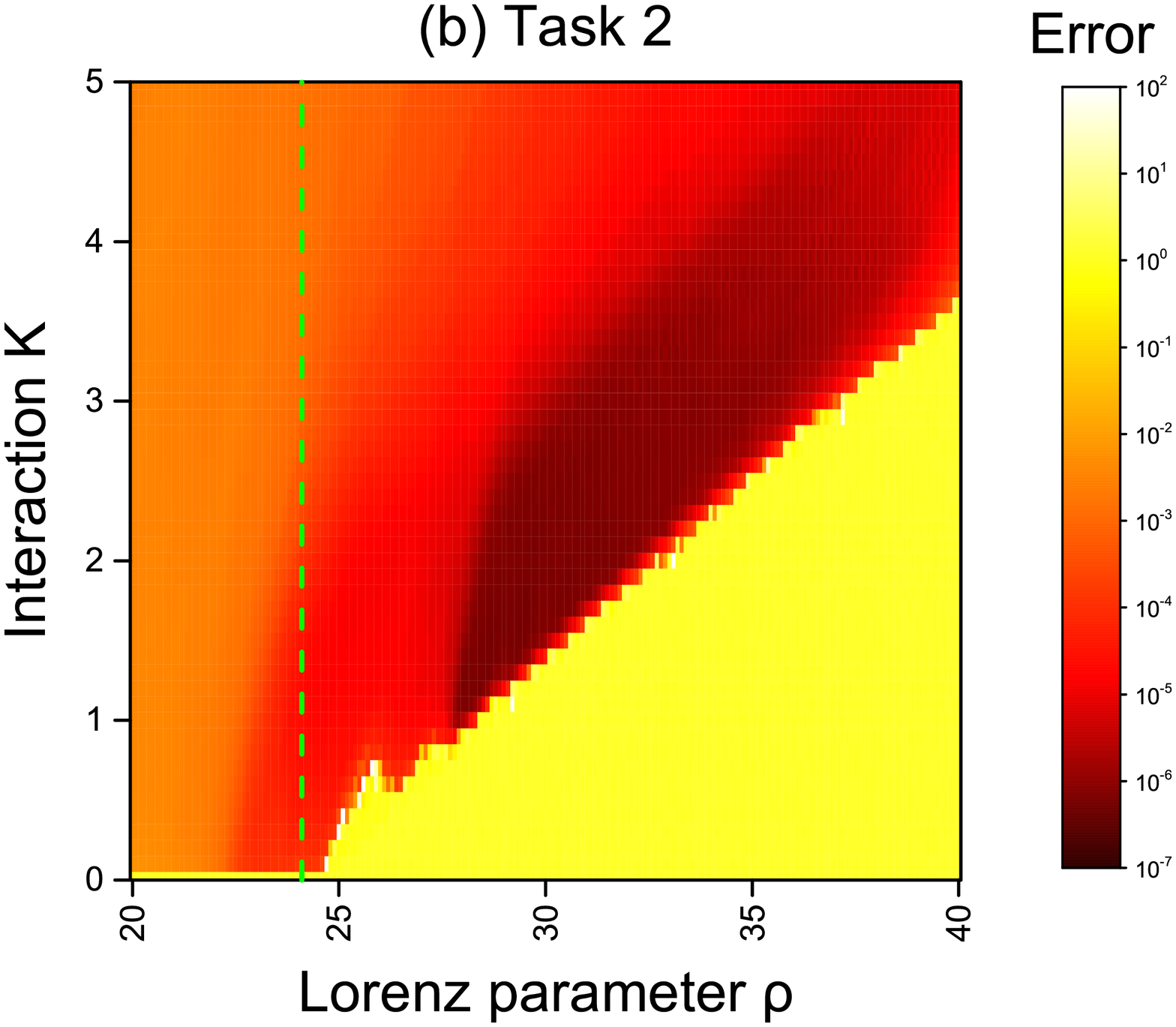} &
		\includegraphics[width=.35\textwidth]{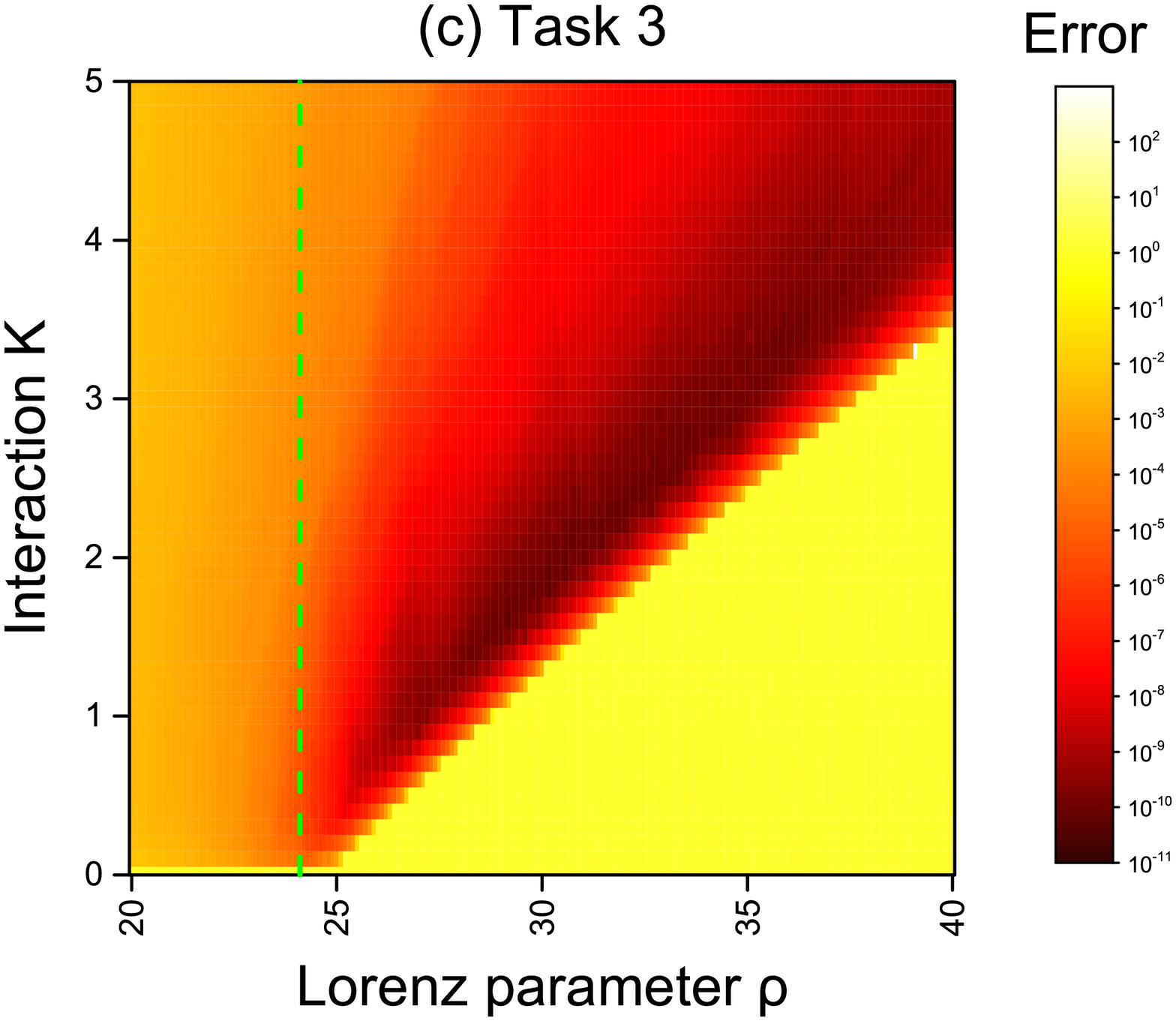} &
	\end{tabular}
	\caption{Test errors according to the coupling $K$ and lorenz system parameter $\rho$. (a) inferring a missing variable of the Rössler system. (b) inferring a missing variable of the Chua's circuit. (c) filtering the Mackey-Glass equation}
\end{figure}

Figure 2 depicts the errors in three tasks with respect the parameters $K$ and $\rho$. One can see in each task that minimum 
error occurs along a diagonal line. The line forms a clear border across 
which the error jumps from the low error regime (red) to high error regime (yellow). It should be noted 
that the three lines are identical and the same as the aligned critical points in Figure 1.(a) where the explosive death of the nodes 
occurs. This assures that the computational performance of the reservoirs is maximized near the first 
order phase transition. 

\subsection{Information capacity of regular and chaotic reservoirs}

In the previous work \cite{CP}, a reservoir that consists of regular phase oscillators was presented as
\begin{equation} \label{eq7}
\theta_i' = \omega_i + \frac{\lambda \vert w_i \vert}{k_i}\sum_{j=1}^{N}A_{ij}\sin(\theta_j-\theta_i), \quad i = 1,\cdots , N,
\end{equation}
where $\lambda$ is the coupling strength of oscillators and $A_{ij}$ is the entry of the adjacency matrix of the network. Here $A_{ij}=1$ if $i\neq j$, otherwise 0. The model in Eq. (7) is known to have a simultaneous synchronization at a certain coupling value $\lambda$ \cite{ZH20}.  Being used as a reservoir, it shows great performance improvement across such critical point. 
 

This section investigates how chaos enhances a criticality in RC. We compare the performance of the forementioned two critical reservoirs, chaotic one in Eq. (1) and regular one in Eq. (7), when both are being poised at the first order phase transition. From here on, we call the former QC(quenched chaos) and the latter ES(explosive synchronization).
When comparing the performace of these reservoirs, it is necessary to consider that the number of equations required to implement a single node is different: if they have the same number of nodes, the computational cost for the reservoir of Eq.(1) is greater than that of the reservoir of Eq. (7), rougly, by a factor of three. 
\begin{figure}[htb]
	\centering
	\begin{tabular}{@{}cccc@{}}
		\includegraphics[width=.35\textwidth]{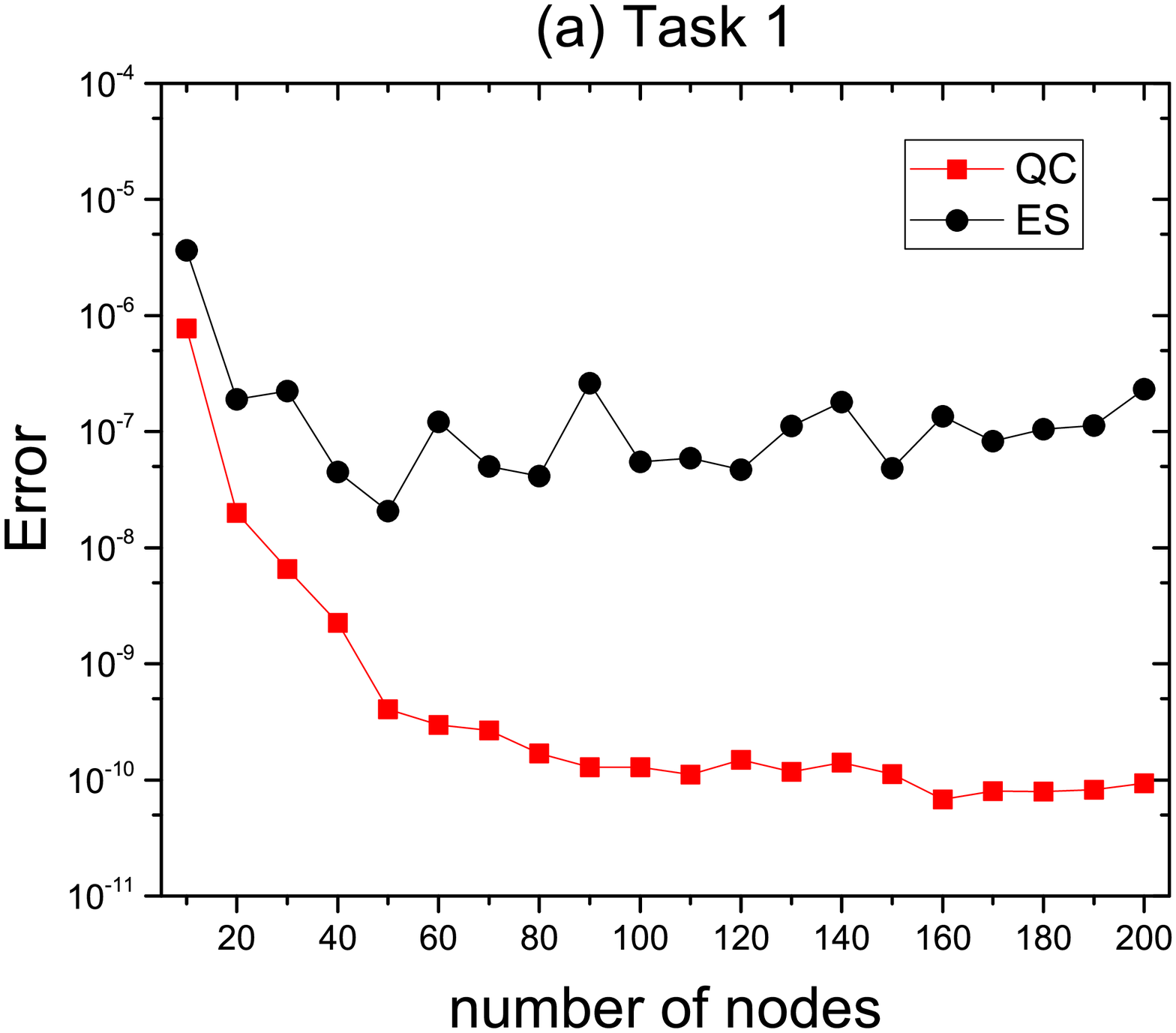} &
		\includegraphics[width=.35\textwidth]{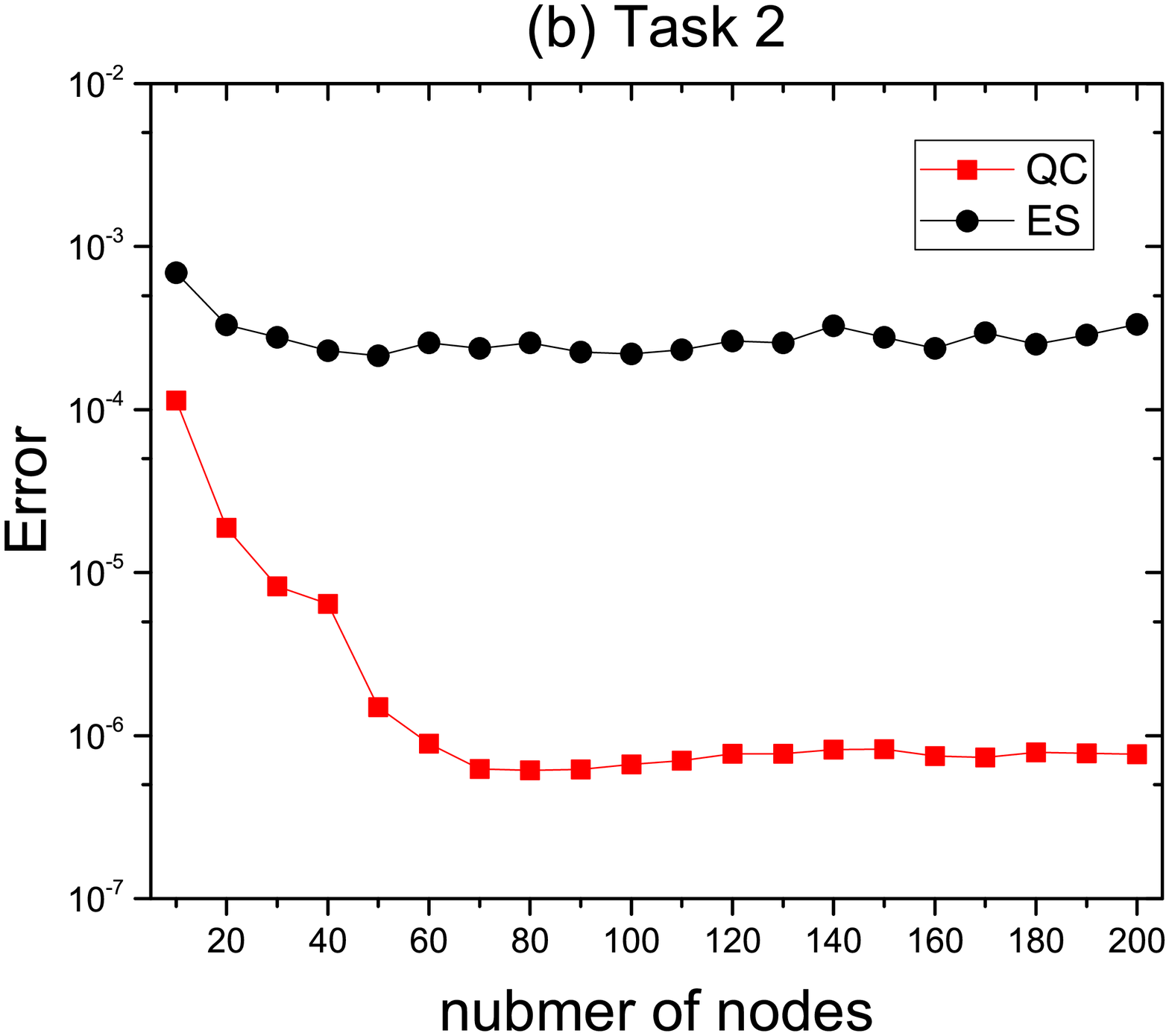} &
		\includegraphics[width=.35\textwidth]{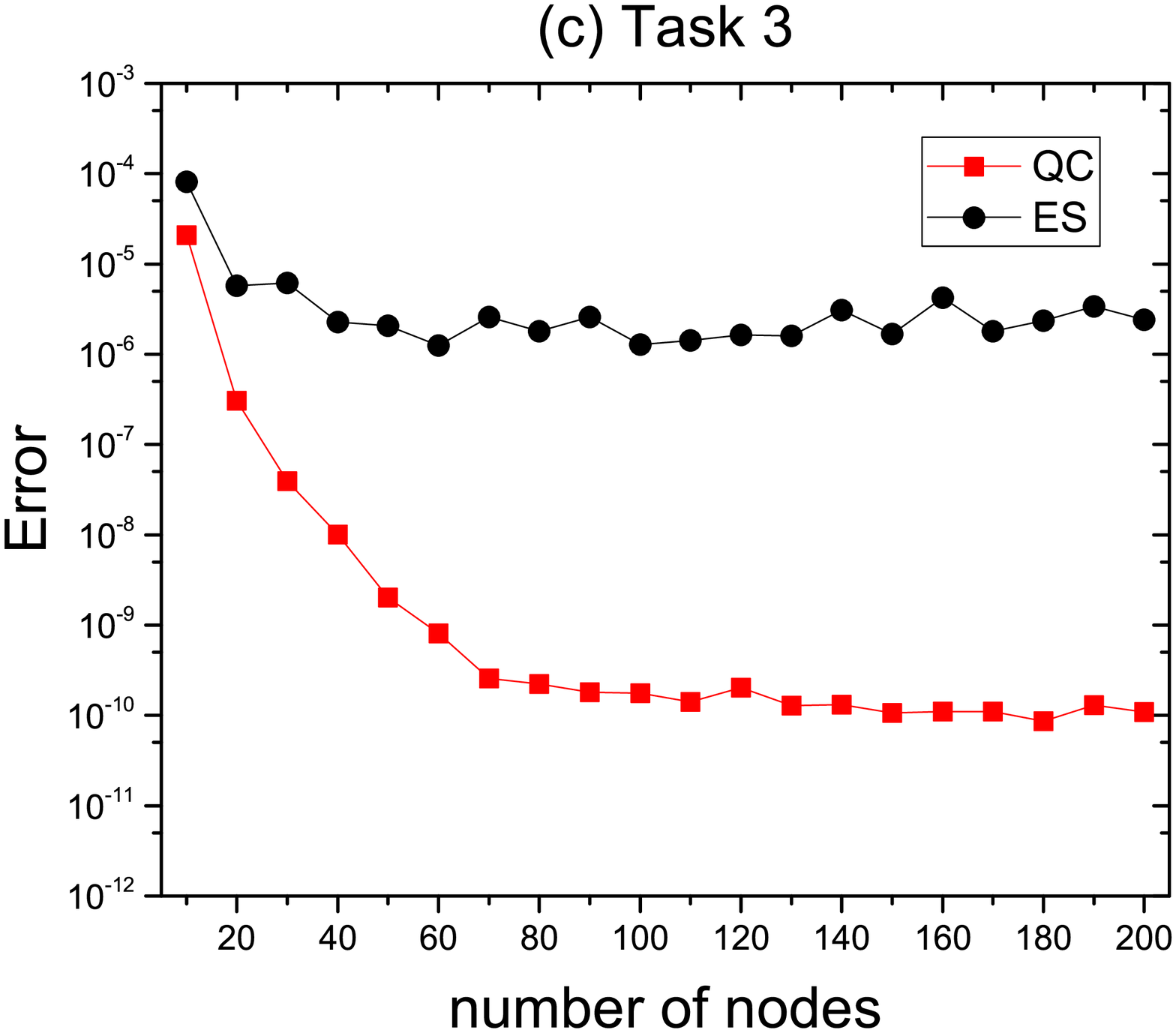} &
	\end{tabular}
	\caption{Comparison of test errors of QC and ES with respect to number of nodes. (a) inferring a missing variable of the Rössler system. (b) inferring a missing variable of the Chua's circuit. (c) filtering the Mackey-Glass equation }
\end{figure}


Figure 3(a) to (c) depict the errors of QC and ES in task 1 to 3, respectively, according to the number of nodes used for the reservoirs.  It is observed that the errors continuously decrease and reaches the minimum  at 150 nodes or less. In all three tasks, the error of QC is at least 1000 times smaller than that of ES. This indicates that QC excels by far ES, even when considering the forementioned difference in computation complexity between two reservoirs.

One of possible explanations on superiority of the chaotic reservoir is that the computing capability of 
critical reservoirs may depend on the collapsed dimension of attractors of reservoirs across the critical 
point. That is, the effect of criticality on computing performance may be related to how much reduction 
occurs in the dimension of the synchronization manifold at the phase transition. One can guess that the 
collapsed dimension of Eq. (1) at the explosive death is much greater than that of Eq. (7), from the fact 
that an attractor of a single Lorenz system has a greater Hausedorff dimension($\sim$2.06), compared to one 
dimensional attractor of a phase oscillator in Eq. (7). Computing the dimension of an 
attractor of a large coupled chaotic system is, however, extremely time-consuming and not practical.

To overcome such difficulty in analyzing reservoir's internal structure, we rather adopt a meaure that focuses on external functional capacity of systems. Here we use the total information capacity which is developed to compute the capacity of any input driven dynamical systems \cite{IC}. The total information capacity, roughly put,  is defined as the assesment of reconstructing the set of orthonormal functions which is a basis of the fading memory Hilbert space.   We refer the reader to \cite{IC} for more details.

\begin{figure}[htb]
	\centering
	\begin{tabular}{@{}cccc@{}}
		\includegraphics[width=.55\textwidth]{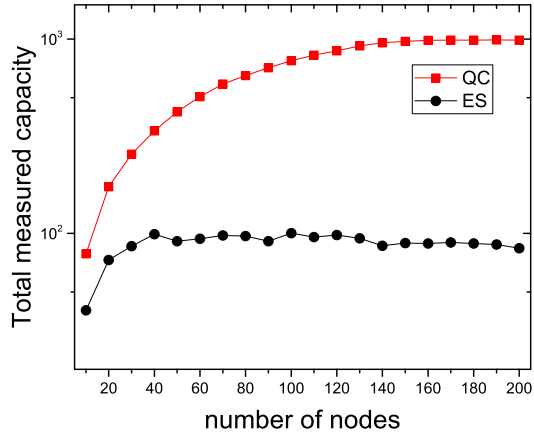}
	\end{tabular}
	\caption{Total information capacity of QC and ES with respect to  number of oscillators. The most of the parameters for evaluation are from \cite{IC}.}
\end{figure}

Figure 4 compares the total information capacity of QC and ES. We confirm that the capacity of QC continuously increases even near 150 nodes then decrease in tendency, while ES only increases till about 50 nodes, which agrees with the results of the numerical tasks in Figure 3.


\section{Discussion}

In this work, we showed that the coupled chaotic systems can be used for efficient reservoir computing. The chaotic reservoirs can create a large criticality at the first order phase transition to create a ground state for computation. It notices in several computing tasks that the chaotic reservoirs excel the regular reservoirs, which is also confirmed from comparing their information capacity.

The results imply that using chaotic nodes is more beneficial in constructing reservoirs. This finding is
important in several aspects. First of all, chaos is widely observed in neuronal systems, both
experimentally and theoretically \cite{AI}. We confirmed that such ubiquity of chaos can be justified from the
perspective of computing performance. That is, as long as it is properly quenched in the critical regime, chaos is an
goal worth pursuing rather than an undesirable state to be avoided. Chaos computing is the paradigm
that exploits the controlled richness of nonlinear dynamics to do flexible computations. This work shows
another theoretical direction of chaos computing different from the approach using chaotic elements to
emulate different logic gates \cite{SI1, SI2}. Basic understanding of a role of criticality in regular and chaotic reservoirs can be expected to shed light on how information is processed in quenched coupled nonlinear systems, potentially leading to proposition of a broad range of reservoirs.

\section*{Acknowledgment}
This work was supported by the Ministry of Education of the Republic of Korea and the National Research Foundation of Korea (NRF-2017R1D1A1B04032921). The funder had no role in study design, data collection and analysis, decision to publish, or preparation of the manuscript.

\bibliographystyle{plain}
\bibliography{QC_references}

\begin{thebibliography}{10}

\bibitem{AI}
Kazuyuki Aihara.
\newblock Chaotic oscillations and bifurcations in squid giant axons.
\newblock {\em Chaos}, pages 257--269, 1986.

\bibitem{RCC1}
Agnesa Babloyantz and Carlos Louren{\c{c}}o.
\newblock Computation with chaos: A paradigm for cortical activity.
\newblock {\em Proceedings of the National Academy of Sciences},
  91(19):9027--9031, 1994.

\bibitem{EC18}
John~M Beggs and Dietmar Plenz.
\newblock Neuronal avalanches in neocortical circuits.
\newblock {\em Journal of neuroscience}, 23(35):11167--11177, 2003.

\bibitem{EC19}
John~M Beggs and Nicholas Timme.
\newblock Being critical of criticality in the brain.
\newblock {\em Frontiers in physiology}, 3:163, 2012.

\bibitem{EC20}
Maria Botcharova, Simon~F Farmer, and Luc Berthouze.
\newblock Markers of criticality in phase synchronization.
\newblock {\em Frontiers in systems neuroscience}, 8:176, 2014.

\bibitem{TK}
Martin Casdagli.
\newblock Nonlinear prediction of chaotic time series.
\newblock {\em Physica D: Nonlinear Phenomena}, 35(3):335--356, 1989.

\bibitem{CP}
Jaesung Choi and Pilwon Kim.
\newblock Critical neuromorphic computing based on explosive synchronization.
\newblock {\em Chaos: An Interdisciplinary Journal of Nonlinear Science},
  29(4):043110, 2019.

\bibitem{AD5}
Michael~F Crowley and Irving~R Epstein.
\newblock Experimental and theoretical studies of a coupled chemical
  oscillator: phase death, multistability and in-phase and out-of-phase
  entrainment.
\newblock {\em The Journal of Physical Chemistry}, 93(6):2496--2502, 1989.

\bibitem{IC}
Joni Dambre, David Verstraeten, Benjamin Schrauwen, and Serge Massar.
\newblock Information processing capacity of dynamical systems.
\newblock {\em Scientific reports}, 2:514, 2012.

\bibitem{EC21}
Bruno Del~Papa, Viola Priesemann, and Jochen Triesch.
\newblock Criticality meets learning: Criticality signatures in a
  self-organizing recurrent neural network.
\newblock {\em PloS one}, 12(5):e0178683, 2017.

\bibitem{AD7}
Milos Dolnik and Irving~R Epstein.
\newblock Coupled chaotic chemical oscillators.
\newblock {\em Physical Review E}, 54(4):3361, 1996.

\bibitem{DU20}
Chao Du, Fuxi Cai, Mohammed~A Zidan, Wen Ma, Seung~Hwan Lee, and Wei~D Lu.
\newblock Reservoir computing using dynamic memristors for temporal information
  processing.
\newblock {\em Nature communications}, 8(1):2204, 2017.

\bibitem{AD1}
GB~Ermentrout and N~Kopell.
\newblock Oscillator death in systems of coupled neural oscillators.
\newblock {\em SIAM Journal on Applied Mathematics}, 50(1):125--146, 1990.

\bibitem{GOUD}
Alireza Goudarzi and Christof Teuscher.
\newblock Reservoir computing: Quo vadis?
\newblock In {\em Proceedings of the 3rd ACM International Conference on
  Nanoscale Computing and Communication}, page~13. ACM, 2016.

\bibitem{AD11}
Ramon Herrero, M~Figueras, J~Rius, F~Pi, and G~Orriols.
\newblock Experimental observation of the amplitude death effect in two coupled
  nonlinear oscillators.
\newblock {\em Physical review letters}, 84(23):5312, 2000.

\bibitem{EC22}
Herbert Jaeger and Harald Haas.
\newblock Harnessing nonlinearity: Predicting chaotic systems and saving energy
  in wireless communication.
\newblock {\em science}, 304(5667):78--80, 2004.

\bibitem{RCC6}
Johannes~H. Jensen and Gunnar Tufte.
\newblock Reservoir computing with a chaotic circuit.
\newblock {\em The 2019 Conference on Artificial Life}, (29):222--229, 2017.

\bibitem{EC60}
Chris~G Langton.
\newblock Computation at the edge of chaos: phase transitions and emergent
  computation.
\newblock {\em Physica D: Nonlinear Phenomena}, 42(1-3):12--37, 1990.

\bibitem{EC27}
Robert Legenstein and Wolfgang Maass.
\newblock Edge of chaos and prediction of computational performance for neural
  circuit models.
\newblock {\em Neural Networks}, 20(3):323--334, 2007.

\bibitem{ED4}
I~Leyva, R~Sevilla-Escoboza, JM~Buld{\'u}, I~Sendina-Nadal,
  J~G{\'o}mez-Garde{\~n}es, A~Arenas, Y~Moreno, S~G{\'o}mez,
  R~Jaimes-Re{\'a}tegui, and S~Boccaletti.
\newblock Explosive first-order transition to synchrony in networked chaotic
  oscillators.
\newblock {\em Physical review letters}, 108(16):168702, 2012.

\bibitem{RCC2}
Carlos Louren{\c{c}}o.
\newblock Attention-locked computation with chaotic neural nets.
\newblock {\em International Journal of Bifurcation and Chaos},
  14(02):737--760, 2004.

\bibitem{RCC3}
Carlos Louren{\c{c}}o.
\newblock Dynamical reservoir properties as network effects.
\newblock In {\em ESANN}, pages 503--508, 2006.

\bibitem{RCC4}
Carlos Louren{\c{c}}o.
\newblock Dynamical computation reservoir emerging within a biological model
  network.
\newblock {\em Neurocomputing}, 70(7-9):1177--1185, 2007.

\bibitem{RCC5}
Carlos Louren{\c{c}}o.
\newblock Structured reservoir computing with spatiotemporal chaotic
  attractors.
\newblock In {\em ESANN}, pages 501--506, 2007.

\bibitem{EC75}
Julian~F Miller and Keith Downing.
\newblock Evolution in materio: Looking beyond the silicon box.
\newblock In {\em Proceedings 2002 NASA/DoD Conference on Evolvable Hardware},
  pages 167--176. IEEE, 2002.

\bibitem{EC1}
Miguel~A Munoz.
\newblock Colloquium: Criticality and dynamical scaling in living systems.
\newblock {\em Reviews of Modern Physics}, 90(3):031001, 2018.

\bibitem{AD2}
I~Ozden, S~Venkataramani, MA~Long, BW~Connors, and AV~Nurmikko.
\newblock Strong coupling of nonlinear electronic and biological oscillators:
  reaching the “amplitude death” regime.
\newblock {\em Physical review letters}, 93(15):158102, 2004.

\bibitem{ADO2}
V~Resmi, G~Ambika, RE~Amritkar, and G~Rangarajan.
\newblock Amplitude death in complex networks induced by environment.
\newblock {\em Physical Review E}, 85(4):046211, 2012.

\bibitem{ADO1}
Amit Sharma and Manish~Dev Shrimali.
\newblock Amplitude death with mean-field diffusion.
\newblock {\em Physical Review E}, 85(5):057204, 2012.

\bibitem{SI1}
Sudeshna Sinha and William~L Ditto.
\newblock Dynamics based computation.
\newblock {\em physical review Letters}, 81(10):2156, 1998.

\bibitem{SI2}
Sudeshna Sinha and William~L Ditto.
\newblock Computing with distributed chaos.
\newblock {\em Physical Review E}, 60(1):363, 1999.

\bibitem{R1}
Gouhei Tanaka, Toshiyuki Yamane, Jean~Benoit H{\'e}roux, Ryosho Nakane, Naoki
  Kanazawa, Seiji Takeda, Hidetoshi Numata, Daiju Nakano, and Akira Hirose.
\newblock Recent advances in physical reservoir computing: a review.
\newblock {\em Neural Networks}, 2019.

\bibitem{ED3}
Umesh~Kumar Verma, Amit Sharma, Neeraj~Kumar Kamal, J{\"u}rgen Kurths, and
  Manish~Dev Shrimali.
\newblock Explosive death induced by mean--field diffusion in identical
  oscillators.
\newblock {\em Scientific reports}, 7(1):7936, 2017.

\bibitem{ED2}
Umesh~Kumar Verma, Amit Sharma, Neeraj~Kumar Kamal, and Manish~Dev Shrimali.
\newblock First order transition to oscillation death through an environment.
\newblock {\em Physics Letters A}, 382(32):2122--2126, 2018.

\bibitem{AD12}
Ming-Dar Wei and Jau-Ching Lun.
\newblock Amplitude death in coupled chaotic solid-state lasers with
  cavity-configuration-dependent instabilities.
\newblock {\em Applied Physics Letters}, 91(6):061121, 2007.

\bibitem{ZH20}
Xiyun Zhang, Xin Hu, J~Kurths, and Zonghua Liu.
\newblock Explosive synchronization in a general complex network.
\newblock {\em Physical Review E}, 88(1):010802, 2013.

\bibitem{ED1}
Nannan Zhao, Zhongkui Sun, Xiaoli Yang, and Wei Xu.
\newblock Explosive death of conjugate coupled van der pol oscillators on
  networks.
\newblock {\em Physical Review E}, 97(6):062203, 2018.

\end{thebibliography}

\end{document}